\documentclass[conference]{IEEEtran}
\IEEEoverridecommandlockouts
\usepackage{multirow}
\usepackage{cite}
\usepackage{amsmath,amssymb,amsfonts}
\usepackage{algorithmic}
\usepackage{graphicx}
\usepackage{textcomp}
\usepackage{xcolor}
\def\BibTeX{{\rm B\kern-.05em{\sc i\kern-.025em b}\kern-.08em
    T\kern-.1667em\lower.7ex\hbox{E}\kern-.125emX}}
\begin{document}
\title{Robust Cross-Etiology and Speaker-Independent Dysarthric Speech Recognition\\
}

\author{\IEEEauthorblockN{Satwinder Singh}
\IEEEauthorblockA{ 
\textit{University of Auckland}\\
 Auckland, New Zealand\\
satwinder.singh@auckland.ac.nz}
\and
\IEEEauthorblockN{Qianli Wang}
\IEEEauthorblockA{
\textit{University of Auckland}\\
 Auckland, New Zealand\\
qwan121@aucklanduni.ac.nz}
\and
\IEEEauthorblockN{Zihan Zhong}
\IEEEauthorblockA{
\textit{University of Auckland}\\
 Auckland, New Zealand\\
zzho680@aucklanduni.ac.nz}
\and
\IEEEauthorblockN{Clarion Mendes}
\IEEEauthorblockA{
\textit{University of Illinois}\\
Champaign, Illinois, USA \\
cmendes2@illinois.edu}
\and
\IEEEauthorblockN{Mark Hasegawa-Johnson}
\IEEEauthorblockA{
\textit{University of Illinois}\\
Champaign, Illinois, USA \\
jhasegaw@illinois.edu}
\and
\IEEEauthorblockN{Waleed Abdulla}
\IEEEauthorblockA{
\textit{University of Auckland}\\
Auckland, New Zealand\\
w.abdulla@auckland.ac.nz}
\and 
\IEEEauthorblockN{Seyed Reza Shahamiri}
\IEEEauthorblockA{
\textit{University of Auckland}\\
Auckland, New Zealand\\
admin@rezanet.com}
}

\maketitle

\begin{abstract}
In this paper, we present a speaker-independent dysarthric speech recognition system, with a focus on evaluating the recently released Speech Accessibility Project (SAP-1005) dataset, which includes speech data from individuals with Parkinson's disease (PD). Despite the growing body of research in dysarthric speech recognition, many existing systems are speaker-dependent and adaptive, limiting their generalizability across different speakers and etiologies. Our primary objective is to develop a robust speaker-independent model capable of accurately recognizing dysarthric speech, irrespective of the speaker. Additionally, as a secondary objective, we aim to test the cross-etiology performance of our model by evaluating it on the TORGO dataset, which contains speech samples from individuals with cerebral palsy (CP) and amyotrophic lateral sclerosis (ALS). By leveraging the Whisper model, our speaker-independent system achieved a CER of 6.99\% and a WER of 10.71\% on the SAP-1005 dataset. Further, in cross-etiology settings, we achieved a CER of 25.08\% and a WER of 39.56\% on the TORGO dataset. These results highlight the potential of our approach to generalize across unseen speakers and different etiologies of dysarthria.
\end{abstract}

\begin{IEEEkeywords}
dysarthria, whisper, ASR, speech accessibility project.
\end{IEEEkeywords}

\section{Introduction}
Automatic Speech recognition (ASR) for individuals with neurological disorders is a complex and challenging task, largely due to the significant variability in speech patterns caused by different underlying conditions \cite{deller1991use}. Dysarthria, a common symptom of several neurological disorders such as PD \cite{moro2019study}, CP \cite{whitehill2000speech}, and ALS \cite{makkonen2018speech}, often results in unintelligible speech that standard ASR systems struggle to accurately transcribe. Further, most existing systems for dysarthric speech recognition are speaker-dependent \cite{raghavendra2001investigation, marini2021optimising, takashima2020two, rudzicz2007comparing} or speaker-adaptive  \cite{shahamiri2023dysarthric, wang2024enhancing}, meaning they require a significant amount of personalized data from each speaker for training. This limitation reduces the generalizability of these systems to new unseen speakers, making them less effective in real-world applications. 
\par
Despite significant advancements in ASR for typical speech \cite{baevski2020wav2vec, schneider2019wav2vec, radford2023robust,singh2022improved, singh2023novel}, developing speaker-independent ASR system for atypical speech remains a challenging area of research. These challenges arise primarily from two factors. Firstly, atypical speech often exhibits significant inter- and intra-speaker variability, atypical pronunciation, and fluctuations in pitch and volume, which hinder the accuracy of current systems \cite{shahamiri2021speech}. Secondly, the lack of large, diverse datasets makes it difficult to train robust speaker-independent systems that can generalize well across different speakers \cite{hernandez2020prosody}.
\par
To address the scarcity of data, various approaches have been explored, including data augmentation \cite{vachhani2018data, mariya2023data, jiao2018simulating, xiong2019phonetic} and speaker-adaptive systems \cite{shahamiri2023dysarthric, wang2024enhancing} that require less data for the target speaker. Recently, researchers proposed a Transformer based on wav2vec 2.0  \cite{baskar2022speaker, hernandez2022cross, javanmardi2024exploring,steinmetz2023transfer} and Whisper model \cite{rathod2023noise, rathod2023transfer}, demonstrating improved performance compared to existing systems. Further, in our previous work \cite{singh2024comprehensive}, we evaluated different versions (English-only, multilingual) and sizes (tiny, small, base, medium) of the Whisper model. Our findings suggested that the medium-sized multilingual Whisper model outperformed other models on the TORGO \cite{rudzicz2012torgo} and UASpeech \cite{kim2008dysarthric} datasets. However, the limited size and diversity of these datasets, particularly in terms of speaker variety and text prompt types (mostly isolated words), posed challenges in building a truly speaker-independent system.
\par
Recently, a collaboration between the University of Illinois and major technology companies led to the development of the Speech Accessibility Project (SAP) \cite{omalley2024uiuc}. SAP offers a unique opportunity to train ASR systems on real-world dysarthric speech, to enhance communication for individuals with speech impairments. The partial release of the SAP-1005 dataset provides a relatively large and diverse set of speakers with PD \cite{zhen2024fine}. Initial work with SAP-1005 \cite{zhen2024fine}, using fine-tuning with wav2vec 2.0, achieved an impressive WER of 26.92\%. While these results are promising, SAP-1005 still requires further evaluation. In this study, we aim to extend this work by evaluating SAP-1005 in a speaker- and severity-independent manner, testing the Whisper model across various sentence categories within SAP-1005. Additionally, by evaluating the fine-tuned Whisper model on the TORGO dataset, we seek to assess the transferability of learned features between different neurological conditions. These findings will contribute to our understanding of cross-etiology transfer learning in ASR, potentially guiding future research toward creating more adaptable and inclusive speech recognition technologies for individuals with diverse speech impairments.
\par
Our experimental analysis demonstrates that the Whisper model produces impressive results in speaker-independent settings on the SAP-1005 dataset, with an overall CER of 6.99\% and WER of 10.71\%. This marks a significant relative improvement of 60.24\% in WER compared to the results reported by Zheng et al. \cite{zhen2024fine} with wav2vec 2.0. We present our findings on the dev\_unshared set, which is distinct from the train and validation sets, as the test set has not yet been made publicly available. Further, we also show that the Whisper model solely fine-tuned on SAP-1005, generalizes well in cross-etiology setting on the TORGO dataset achieving CER of 25.08\% and WER of 39.56\%.
With this paper, we made the following key contributions:
\begin{itemize}
    \item Comprehensive evaluation of the SAP-1005 dataset, offering insights into its utility for speaker-independent dysarthric speech recognition research.
    \item Evaluate the Whisper model's ability to generalize across different types of dysarthria (hypokinetic 	$\rightarrow$ spastic, flaccid, or ataxic) present in SAP-1005 and TORGO datasets.
\end{itemize}

\section{Related Work}
Most existing work on dysarthric speech has primarily focused on speaker-dependent \cite{raghavendra2001investigation, marini2021optimising, takashima2020two, rudzicz2007comparing} and adaptive ASR systems \cite{shahamiri2023dysarthric, wang2024enhancing}, with relatively little attention given to speaker-independent systems. This gap in the literature is largely due to the scarcity of large, labeled datasets for dysarthric speech. As a result, current speaker-independent systems often fail to produce useful results and struggle to generalize across different etiologies.
\par
Yilmaz et al. \cite{yilmaz2019articulatory} presented a speaker-independent ASR system for dysarthric speech in Dutch and Flemish, utilizing a time-frequency convolutional neural network (TFCNN) to compare the performance of bottleneck and articulatory features. Additionally, self-supervised cross-lingual models such as wav2vec 2.0, Hubert, and XLSR have been employed for speaker-independent (SI) dysarthric speech recognition \cite{hernandez2022cross}, demonstrating that the multilingual XLSR model achieved the best WER across UASpeech, EasyCall, and PC-GITA datasets. Bhat et al. \cite{bhat2022improved} introduced a two-stage data augmentation approach that leveraged both static (e.g., speed, tempo, volume, and reverse deep autoencoder) and dynamic (e.g., dysarthric SpecAugment \cite{park2019specaugment}) augmentation techniques. Their experiments on the UASpeech dataset revealed that combining all data augmentation methods led to a 20.6\% WER. However, this work was limited to isolated words. 
\par
Zheng et al. \cite{zhen2024fine} presented the first work on the SAP-1005 dataset, introducing a series of fine-tuning strategies aimed at enhancing dysarthric and dysphonic speech recognition. Their study explored several methods, such as speaker clustering, severity-dependent models, weighted fine-tuning, and multi-task learning. Among these, the multi-task learning approach, which integrated ASR with severity estimation, yielded the most promising results. This approach significantly reduced WER by 37.62\% and 26.97\% when compared to models fine-tuned on 100 hours and 960 hours of LibriSpeech, respectively.

\section{Data}
\subsection{Speech Accessibility Project}
The Speech Accessibility Project (SAP) \cite{omalley2024uiuc}  is dedicated to collecting a broad array of speech samples to enhance speech-related technologies for individuals with various speech disabilities. Led by a team at the University of Illinois Urbana-Champaign, SAP is a collaborative effort involving major technology companies including Apple, Google, Microsoft, Meta, and Amazon.
The dataset, partially released on 2023-10-05, for six participating institutes, contains data from 253 individuals diagnosed with PD. This dataset, named SAP-1005 \cite{zhen2024fine}, includes 190 speakers in the train set, 21 in the dev set, and 42 in the test set. The dev and test sets are further divided into shared and unshared portions. The shared portion contains some text prompts that overlap with the train set, while the unshared portion does not include any common text prompts. It is important to note that although the shared sections of the dev and test sets include some of the same text prompts, these prompts are recorded by entirely different speakers, making SAP-1005 a speaker-independent dataset. 
\par
While the dataset is available for research purposes to other researchers, the test set is reserved for an upcoming competition and is not publicly available. Therefore, we utilized only the train and dev sets, which encompass speech data from 211 individuals, representing a diverse group of 119 male and 92 female speakers. This dataset comprises a total of 174.79 hours of speech. As per \cite{zhen2024fine}, the severity classes were determined objectively based on the CER of a wav2vec 2.0 model fine-tuned on the LibriSpeech 960h dataset. Speakers with a CER of less than 10\% were classified as Very Low (VL), between 10\% and 20\% as Low (L), between 20\% and 40\% as Median (M), and greater than 40\% as High (H). Additionally, the dataset features three distinct types of sentence categories: digital assistant commands (67\%), novel sentences (22\%), and spontaneous speech prompts (11\%). Digital Assistant Commands (DAC) include simple, task-oriented commands similar to those used with digital assistants like Apple Siri or Amazon Alexa, providing practical insights into everyday usage scenarios. Novel Sentences (NS) are read speech samples, offering lexical diversity.  Lastly, the Spontaneous Speech Prompts (SSP) category contains conversational, spontaneous speech, reflecting more natural, unstructured communication patterns.

\subsection{TORGO}
The TORGO dataset \cite{rudzicz2012torgo}, developed by the University of Toronto, offers labeled speech data specifically from individuals with dysarthria resulting from CP and ALS. It comprises approximately 23 hours of audio recordings, featuring speech from 8 individuals with dysarthria (5 males and 3 females) and 7 control speakers without speech impairments (4 males and 3 females). The dataset offers a mix of both isolated words and continuous speech utterances. The dysarthric participants are classified into severity categories—severe, moderate/severe, moderate, and mild. For this work, we only utilize continuous speech utterances from the TORGO dataset.

\section{Our Approach}
\subsection{The Whisper Model}
We adopt a pre-trained Whisper model \cite{radford2023robust} for our approach. The Whisper model employs a encoder-decoder based transformer architecture to perform ASR by mapping audio inputs to text outputs. The process begins with the conversion of an audio waveform \( x(t) \) into a spectrogram \( S(x(t)) \), which represents the time-frequency characteristics of the speech signal. The spectrogram \( S \) is then fed into the model's encoder, a series of transformer layers, each defined by a self-attention mechanism:

\[
\text{Self-Attn}(Q, K, V) = \text{softmax}\left(\frac{QK^T}{\sqrt{d_k}}\right)V
\]

where \( Q \), \( K \), and \( V \) are the query, key, and value matrices, and \( d_k \) is the dimensionality of the key vectors. This mechanism enables the model to capture contextual relationships across different parts of the input sequence.

The encoded representations \( H \) from the encoder are then passed to the decoder, which generates the text output by predicting a sequence of tokens \( y_1, y_2, \dots, y_T \) corresponding to the input audio. The decoder operates autoregressively, where the probability of the next token \( y_t \) is conditioned on the previous tokens and the encoded input:
\[P(y_t \mid y_1, y_2, \dots, y_{t-1}, H)\]
Beam search \cite{graves2012sequence} is employed during decoding to explore multiple potential sequences \( Y \), optimizing the output by selecting the sequence that maximizes the overall probability: 
\[P(Y \mid H)\]

\subsection{Methodology}
Our previous work \cite{singh2024comprehensive} demonstrated that the medium-sized multilingual  Whisper model outperformed other versions when tested with dysarthric speakers. Therefore, in this paper, we conduct all experiments using the medium-sized multilingual Whisper model. During our initial investigations, we observed that the model performed poorly on some speech utterances in the SAP-1005 dataset. Further analysis revealed that the model tends to hallucinate on speech utterances longer than 30 seconds \cite{radford2023robust}, leading to higher error rates. This issue arises because the Whisper model has a receptive field of 30 seconds, beyond which it starts repeating output tokens. To address this problem, we divided the longer speech utterances into 30-second chunks with a 5-second overlap between chunks. Each chunk is decoded individually, and the entire output transcript is then concatenated.  

We fine-tuned the model on the SAP-1005 dataset with an initial learning rate of 1e-5, selecting the best-performing validation checkpoint for testing. Since the official test set for SAP-1005 is reserved for an upcoming competition and is not publicly available, we repurposed the dev\_unshared set as a test set and the dev\_shared set as a validation set. For decoding, we set \texttt{num\_beams=10}, \texttt{no\_repeat\_ngram\_size=3}, and penalized longer outputs with \texttt{length\_penalty=1.0}.

\section{Results and Discussions}
\subsection{Speaker-independent Results on SAP-1005}

The empirical results presented in Table \ref{tab:sap} demonstrate impressive performance with lower error rates on the SAP-1005 dataset. Compared to the initial work by Zheng et al. \cite{zhen2024fine}, our Whisper model achieved a notable overall WER of 10.71\% on the dev\_unshared set, representing a relative improvement of 60.24\%. While this indicates a substantial enhancement over the WER of 26.92\% reported by Zheng et al., it is important to note that the comparison is limited by the fact that we evaluated different portions of the dataset. Nevertheless, these results highlight the potential of Whisper as a strong baseline for speaker-independent ASR for dysarthric speech.

\begin{figure}[b]
    \centering
    \includegraphics[width=\linewidth]{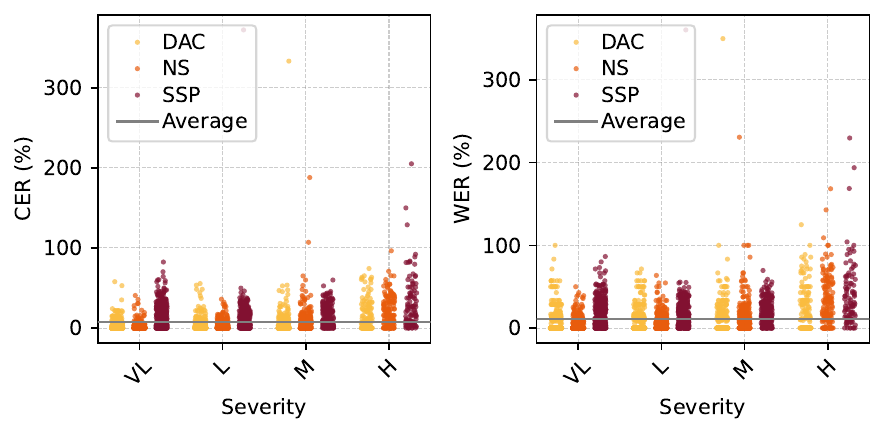}
    \caption{Strip plot showing error rate of each utterance in dev\_unshared set of SAP-1005 across different sentence category and severity level.}
    \label{fig:stripplot}
\end{figure}


\subsubsection{Severity-based Results}
The empirical results, as shown in Table \ref{tab:sap} and Figure \ref{fig:stripplot}, demonstrate strong performance for speakers with very low and low severity dysarthria. The model achieved impressive accuracy, with a CER of 4.73\% and WER of 7.12\% for very low severity, and similar performance for low severity, with a CER of 4.68\% and WER of 7.76\%. These results highlight the model’s effectiveness in handling mild cases of dysarthria. For speakers with median severity, the CER rises to 8.8\% and WER to 13.24\%. The most significant challenge lies in the high severity category, where the model struggles more, yielding a CER of 20.71\% and WER of 30.51\%. Despite this, the overall results underline the model’s robustness for less severe dysarthria, while also pinpointing the difficulties of transcribing speech in more severe cases.


\subsubsection{Category-based Results}
In our initial analysis, we evaluated the model's performance across different sentence categories without considering the severity levels. As shown in Table \ref{tab:sap}, the model produces the lowest WER for DAC and NS. The highest WER is observed for SSP, indicating that the model struggles more with spontaneous speech, which typically includes more naturalistic and unpredictable language patterns.

\begin{table}[]
\centering
\caption{Experimental results on SAP-1005 in terms of severity and sentence category}
\begin{tabular}{lll}
\hline
\textbf{Severity}                     & \textbf{CER (\%)} & \textbf{WER(\%)} \\ \hline
Very Low                          & 4.73     & 7.12    \\
Low                          & 4.68     & 7.76    \\
Median                            & 8.80      & 13.24   \\
High                           & 20.71    & 30.51   \\ \hline\hline
\textbf{Category}                     & \textbf{CER (\%)} & \textbf{WER(\%)} \\ \hline
Digital   Assistant Commands & 3.86     & 7.92    \\
Novel   Sentences            & 4.65     & 8.05    \\
Spontaneous   Speech Prompts & 15.62    & 19.29   \\ \hline
\textbf{Overall average }                     & \textbf{6.99}       & \textbf{10.71}   \\ \hline
\end{tabular}
\label{tab:sap}
\end{table}

\begin{figure}[htbp]
    \centering
    \includegraphics[width=\linewidth]{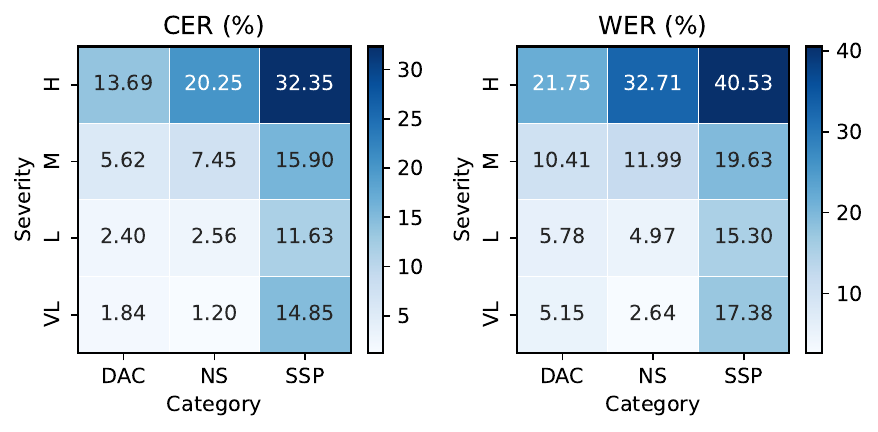}
    \caption{Experimental results across different severity levels and sentence categories in the SAP-1005.}
    \label{fig:severity_category}
\end{figure}

To gain a deeper understanding of the model's performance, we conducted a fine-grained analysis by examining the performance across different severity levels within each sentence category (Figure \ref{fig:severity_category}). The results reveal that, while the model performs relatively well on DAC across all severity levels, there is an increase in error rates as the severity of the speech impairment increases, particularly in SSP. For instance, within the High severity group, the WER for SSP is 30.51\%, compared to just 7.92\% for DAC, illustrating the substantial challenge posed by spontaneous speech in more severe cases. The low performance in SSP can be attributed to two primary factors. First, many SSP utterances in the SAP-1005 dataset are relatively long, often up to 120 seconds. This challenges the Whisper model's limited receptive field and leads to hallucinations despite our attempts to mitigate this by chunking the longer utterances \cite{radford2023robust}. Second, individuals with PD often produce less intelligible speech in spontaneous settings due to the lack of external cues  \cite{venetjoki2006effect}, resulting in disorganized speech with inconsistent pauses, pitch variations, volume fluctuations, and disfluencies. This is evident in Figure \ref{fig:stripplot}, where error rates exceeding 100\% highlight these challenges.

\subsection{Cross-Etiology Results on TORGO}
We evaluate the transferability of the Whisper model by fine-tuning it on speech data from individuals with hypokinetic dysarthria caused by PD and testing it on spastic, flaccid, or ataxic dysarthria caused by CP and ALS using the TORGO dataset. Hypokinetic dysarthria is marked by reduced movement, resulting in soft, imprecise, and monotonous speech patterns with diminished volume \cite{metter1986clinical}. In contrast, spastic dysarthria is defined by hypertonia, producing strained and effortful speech with irregular rhythm, pitch, and loudness \cite{hustad2002spastic}. Flaccid dysarthria, distinguished by muscle weakness, leads to breathy, nasal speech with imprecise articulation and reduced loudness \cite{enderby2013disorders}. Ataxic dysarthria, on the other hand, is characterized by incoordination of the speech muscles, resulting in slurred, irregular, and imprecise speech with disrupted timing and rhythm \cite{kent2000ataxic}. Given the significant differences between these types of dysarthria, the Whisper model demonstrated a surprising degree of generalization achieving a CER of 25.08\% and a WER of 39.56\% as shown in Table \ref{tab:torgo}.

While the performance on the TORGO dataset is lower compared to SAP-1005, the results underscore the Whisper model's capacity to transfer learned features across different types of dysarthria, despite significant variations in speech patterns. This relative success in cross-etiology transfer suggests that the model captures underlying commonalities in dysarthric speech across different neurological conditions. However, further investigation, model refinement, and the inclusion of more diverse training data are necessary to improve generalization across different etiologies.

\begin{table}[]
\caption{Cross-etiology results on TORGO dataset. The model is only fine-tuned on the SAP-1005 dataset. }
\centering
\begin{tabular}{llll}
\hline
\textbf{Intelligibility}                  & \textbf{Speaker} &\textbf{ CER (\%)} & \textbf{WER (\%)} \\ \hline
Severe                           & M04     & 50.66    & 76.4     \\ \hline
\multirow{4}{*}{Moderate/Severe} & F01     & 41.70     & 65.33    \\
                                 & M01     & 31.63    & 51.8     \\
                                 & M02     & 36.03    & 54.64    \\ \cline{2-4} 
                                 & \textbf{Average} & \textbf{36.45}    & \textbf{57.26}    \\ \hline
Moderate                         & M05     & 27.58    & 42.6     \\ \hline
\multirow{4}{*}{Mild}            & F03     & 10.63    & 20.21    \\
                                 & F04     & 1.54     & 3.35     \\
                                 & M03     & 0.84     & 2.12     \\ \cline{2-4} 
                                 & \textbf{Average} &\textbf{ 4.34}     & \textbf{8.56}     \\ \hline
\multicolumn{2}{l}{\textbf{Overall   average}}      & \textbf{25.08}    & \textbf{39.56}    \\ \hline
\end{tabular}
\label{tab:torgo}
\end{table}

\section{Conclusions}
Our study demonstrates the efficacy of the Whisper model in speaker-independent dysarthric speech recognition, achieving significant improvements on the SAP-1005 dataset. Additionally, the model exhibited reasonable generalization in a cross-etiology setting on the TORGO dataset, despite the distinct differences between hypokinetic and spastic, flaccid or ataxic dysarthria. We also observed that the Whisper model demonstrates higher accuracy in recognizing digital assistant commands and novel sentences compared to spontaneous speech prompts. These findings indicate the potential of Whisper in advancing robust ASR systems for dysarthric speech, paving the way for more inclusive and adaptable speech recognition technologies.
\clearpage
\bibliographystyle{IEEEtran}
\bibliography{references}

\end{document}